# Terahertz birefringent gratings for filtering and dispersion compensation

Muhammad Talal Ali Khan[1*], Haisu Li [2*], Yajing Liu[2], Gang-Ding Peng[3], and Shaghik Atakaramians[1]

[1]Shaghik's THz Group, School of Electrical Engineering and Telecommunications, UNSW Sydney, New South Wales 2052, Australia
[2]Key Laboratory of All Optical Network and Advanced Telecommunication Network of EMC, Institute of Lightwave Technology, Beijing Jiaotong University, Beijing 100044, China
* Contributed equally to this work.
m.t.khan@unsw.edu.au

The recent development of the terahertz waveguide makes it an excellent platform for integrating many intriguing functionalities, which offers tremendous potential to build compact and robust terahertz systems. In the context of next-generation high-speed communication links at the terahertz band, engineering the dispersion and birefringence of terahertz waves is essential. Here, we experimentally demonstrate subwavelength birefringent waveguide gratings based on the low-loss cyclic olefin copolymer exploiting micro-machining fabrication techniques. Asymmetric cross-section and periodic-structural modulation along propagation direction are introduced to achieve birefringent THz grating for filtering and dispersion compensation. Because of strong index modulation in the subwavelength fiber, a high negative group velocity dispersion of -188 (-88) ps/mm/THz is achieved at 0.15 THz for $x$-polarization ($y$-polarization), i.e., 7.5 times increase compared to the state-of-the-art reported to date. Such high negative dispersion is realized in a 43 mm grating length, which is less than half of the length reported until now (e.g., 100 mm). Further, the subwavelength fiber grating filters two orthogonal polarization states and exhibits transmission dips with 8.5-dB and 7.5-dB extinction ratios for $x$ and $y$ polarization, respectively. Our experiment demonstrates the feasibility of using polymer-based terahertz gratings as a dispersion compensator in terahertz communications and steering polarized terahertz radiations for polarization-sensitive THz systems.

Over the last few years, terahertz (THz) technology has evolved dramatically due to the successful implementation of interesting applications such as high-speed communication, non-destructive material inspection, biological sensing, security, and many others [1]. These diverse applications require not only sources and detectors for the generation and detection of THz waves but also the waveguide components [2]. Although the development of various THz waveguide structures has progressed quickly between 0.1 THz to 1 THz frequencies range [3, 4], a key challenge is that most of the polymer materials have high absorption losses. Over the past decades, mostly subwavelength THz fibers (propagation mainly in the air cladding) [5-7] or air-core structures with exotic cladding designs (propagation in the air core) [8-11] have been demonstrated as the promising platform to mitigate high transmission losses of polymers. To effectively manipulate THz waves, waveguide-based functional devices including filters, couplers, polarizers, modulators, sensors [12-17] are also particularly important to realize the THz applications like wireless communication systems [16], THz spectroscopy [18], sensing systems [19] and high-resolution imaging [20]. Out of many aforementioned devices, steering polarization states are also crucial for polarization-sensitive THz systems, which can be effectively realized via designing birefringent waveguides [6].

The waveguide-based gratings offer important functionalities such as bandstop/bandpass filtering [12, 21-23] and dispersion manipulation, which have been extensively applied at optic communication band (i.e., the well-known optical fiber gratings) [24], while the demonstrations at THz frequencies are still limited despite such intriguing applications [16, 18-20]. In 2012, a THz notch-type Bragg grating written on a polymer fiber was reported for filtering applications, however, one polarized state was found to be stronger than other polarization states due to fabrication limitations — laser cutting technique caused the asymmetric distribution of etched gratings on circular rod [25]. Another THz waveguide grating has been demonstrated based on a plasmonic two-wire waveguide accompanied by paper grating, yet it only filters the wave perpendicular to the paper grating [26]. Subsequently, Ma *et al*. [16] demonstrated hollow-core THz metallic waveguide Bragg gratings for dispersion compensation in THz communication links. A negative dispersion of -25 ps/THz/mm at 0.14 THz was achieved in a 100 mm long grating device. However, such a circular waveguide is polarization-independent and has weak modulation due to an oversized core. We notice that most of the reported THz gratings' either filter one polarization state or are polarization independent. Recently, motivated by the demands of polarization-sensitive THz systems, we have numerically proposed a THz subwavelength birefringent waveguide grating as a stopband filter [12]. This grating filters two orthogonal polarization states simultaneously and is expected to be a promising candidate for upcoming THz communications. For example, the wireless transmission capacity can be doubled by polarization-division-multiplexing techniques when two orthogonal polarized THz waves propagate simultaneously in a multiple-input multiple-output THz wireless transmission system [27].

In this work, we experimentally demonstrate THz birefringent grating, which filters two polarization states simultaneously and offers large negative dispersion in the shortest length. Our experimental results confirm the filtering of two orthogonal modes at 0.15 THz. The measurements show 8.5-dB and 7.5-dB extinction ratios (ER) with a 5 GHz full-width at half-maximums (FWHM) for x-polarization and y-polarization respectively. We also demonstrate a large negative group velocity dispersion of -188 ps/mm/THz in a 43 mm device length, which is more than 7.5 times higher dispersion in less than half of the length reported to date in THz waveguide-based gratings. Our demonstration shows a promising path for filtering and dispersion compensation in future THz communication links. In this paper, we present the design and the fabrication of THz birefringent grating in Section 2. In Section 3, the characterization setup, transmission characteristics, and group velocity dispersion are investigated. In Section 4, we discuss and conclude the findings.

Figure 1 (a) presents a schematic of the subwavelength birefringent waveguide-based grating structure, where the THz waves propagate along the z-axis (longitudinal direction). The waveguide-based grating is constructed by introducing the structural perturbation periodically along the propagation direction. The geometrical birefringence in rectangular waveguide cross-section supports two orthogonally polarized modes i.e., x-polarization (x-pol.) and y-polarization (y-pol.). As shown in the inset of Fig. 1 (a), the single grating unit consists of large ($C_1$) and small ($C_2$) sized rectangular cells and supports modes with effective refractive indices of $n_{eff1}$ and $n_{eff2}$, respectively. This results in periodic structural modulation due to resonant mode coupling along a longitudinal direction. Such coupling shows a stopband profile in the transmission spectrum, where the Bragg frequency ($f_B$) of the stopband can be estimated by Bragg condition as follows [12]:

$$2f_B(n_{eff1}L_1 + n_{eff2}L_2) = mc \qquad (1)$$

where $m$ is an integer ($m$= 1 in this work), $c$ is the speed of light in vacuum, $L_1$ and $L_2$ are the lengths of the $C_1$ and $C_2$, respectively.

To achieve a target frequency band at 0.15 THz, we design waveguide gratings according to Eq. (1) with appropriate cross-sectional parameters of $d_{1x}$, $d_{1y}$, $d_{2x}$, $d_{2y}$. The grating pitch ($\Lambda$) and total grating length (period number of $N$) are equal to $L_1 + L_2$ and $N \times \Lambda$, respectively. We fabricated two gratings with period numbers 29 and 45. The gratings were manufactured using micromachining techniques. The photographs of the two fabricated samples are shown in Fig. 1 (b), where the total measured lengths are 30 mm (denoted as short) and 43 mm (denoted as long), respectively. The segments were machined on a Kira Super Mill M2 which is a 3-axis precision (2 µm) milling machine using a Fanuc31i controller [28]. The cutting programs used on the Kira were created using SolidCAM CAD/CAM software, which also optimized the cutting parameters. The samples were inspected using an Olympus Laser scanning microscope [29]. The microscope images of short and long samples are shown in Fig. 1 (c). To reduce propagation losses, we select a low-loss cyclic olefin copolymer (COC) material with the tradename of TOPAS® COC 5013L-10 [30]. We use a COC sheet as it is relatively easier to cut the grating using the micromachining process. The thickness of the sheet is 2 mm. We utilize THz time-domain spectroscopy (THz-TDS, see details in Section 3.1) to characterize the properties of the COC. Figure 2 (a) shows the measured (experiment fitting) refractive index and absorption coefficient of COC. The index of refraction is 1.536 at 0.15 THz and constant up to 1 THz. As expected, we observed a very low absorption coefficient (0.0027 1/mm) at 0.15 THz. Based on the material properties, the extracted complex refractive index of COC at 0.15 THz is 1.538+0.0065i, which is used in the Lumerical FDTD simulations [31]. The measured parameters of short and long fabricated samples are summarized in Table 1. We will discuss the impact of fabrication deviations on the performance of grating in Section 3.2.

Table 1. Measured parameters of the fabricated gratings samples

| Parameter | Short grating | Long grating |
|---|---|---|
| $L_1(\mu m)$ | 580 ± 10 | 580 ± 10 |
| $L_2(\mu m)$ | 240 ± 20 | 240 ± 10 |
| $d_{1x}(\mu m)$ | 1340 ± 50 | 1300 ± 20 |
| $d_{2x}(\mu m)$ | 865 ± 40 | 906 ± 30 |
| $d_{1y}(\mu m)$ | 938 ± 40 | 1020 ± 20 |
| $d_{2y}(\mu m)$ | 606 ± 50 | 711 ± 10 |

*Experimental Setup*

We employ a THz-TDS system (Menlo TeraSmart [32]) to characterize the performance of fabricated gratings. Figure 2 (c) shows the image of the transmission characterization setup. The measured bandwidth of the THz-TDS system is from 0.1 THz to 3.5 THz with over 65 dB of dynamic range. The THz pulse is generated by the fiber-coupled THz antenna that is biased and irradiated by an ultrafast femtosecond laser. Note that the THz source and detector are linearly polarized, oriented along the *y*-direction in the setup. We coupled the free-space propagation of THz pulse into the grating samples by employing a pair of TPX lenses (TPX: polymethylpentene lens with a focal length of 50 mm). Before the sample, a pinhole of 1.5 mm diameter is used, to align the grating to the center of the beam. We use a dielectric foam to hold the gratings, as shown in the inset of Fig. 2 (c). We investigate different dielectric foams and, has chosen the foam with the closet refractive index to air, [see Fig. 2(b)], which is expected to exhibit a negligible effect on the transmission properties of the gratings. The measured properties of the foam can be seen in Fig. 2 (b) which shows a refractive index of 1.004 and low losses of 0.003 1/mm in the vicinity of 0.15 THz. It should be noted that the two polarizations are measured by rotating the waveguide grating sample orthogonally. After passing through the sample under test, the beam is recollimated and focused into the detector (THz antenna) by an identical pair of lenses.

*Transmission Measurements*

Figures 3 (a) and 3 (b) show the transmission spectra of the short and long gratings for *x*- and *y*-polarizations, where the solid and dashed curves represent measured and numerical results, respectively. For short grating [Fig. 3(a)], the measured reflective frequencies of *x*- and *y*-polarizations are at 0.152 THz and 0.155 THz, with ERs of 5.8-dB and 5.0-dB, respectively. Note that, we consider averaged transmission outside the stopbands (green dashed line) for an approximate evaluation of experimental ERs for both polarizations. The FWHM of the stopbands are 8 (9) GHz, for *x*-pol. (*y*-pol.). For long grating [Fig. 3(b)], 8.5- and 7.5-dB ERs are observed for *x*-pol. and *y*-pol., respectively, and FWHM of 5 GHz for both polarizations. The measure reflective frequencies of long grating for *x*-pol. (*y*-pol.) are at 0.1504 (0.1528) THz. As expected, increasing the grating length leads to high ER and eventually shows strong modulation strength. This shows that the index modulation of the unit cell of the short grating is weaker than the long grating which results in different FWHMs of two gratings.

In general, there is a good agreement between measured (solid curves in Fig. 3) and numerically calculated (dotted curves in Fig. 3) stopband positions of two gratings. The stopband positions of the short grating are the same as of numerical, while a very small deviation of 0.13 % is observed between the simulation and experiment stopband of the long grating (*x*-pol. only). In terms of ERs and FWHMs, there is a slight mismatch in the measured results when compared to numerical simulations. For example, the *x*- and *y*-polarized experimental ERs of the short grating show mismatch of 2.2-dB and 1.5-dB, respectively. On the other hand, the ERs of the long grating illustrate a much smaller mismatch of 0.5-dB (*x*-pol.) and 0.2-dB (*y*-pol.). Furthermore, the measured FWHM is wider, particular for short grating than the numerically calculated FHWMs, which are 5.5 and 4.5 GHz for short and long gratings, respectively. We attribute the deviations between experiment and simulation to fabrication tolerances. Obviously, the grating's unit cells are identical in the numerical simulations, however, the grating's unit cells of fabricated samples have slight variations (see Table 1). To understand the effect of variations on the filtering characteristics including ER and FWHMs, we numerically study the impact of the two key parameters of the grating unit structure ($C_1$ and $C_2$) [12]. We vary the relative fraction ($L_1/\Lambda$) of $C_1$ in the longitudinal direction and the cross-sectional parameter ($d_x$) along the transverse (cross-sectional) direction. The numerical results suggest the reduction in ER (5-dB) and narrowing FWHM (3 GHz) of stopband for both polarization states due to weak mode coupling if the corresponding relative fraction increases (+ 50 um). This is while the decrease (- 50 um) in relative fraction can widen (5 GHz) the stopband FWHM. Furthermore, we observe low transmission (mostly for short grating) in the experimental measurements compared with simulation. We attribute that to relatively weak coupling from free space, which can be due to slight misalignment (sensitive) of the waveguide grating with an incoming THz beam. To check the effect of misalignment on the coupling efficiency, we numerically estimate the coupling efficiency of the waveguide grating using an input terahertz gaussian beam (beam waist of 1.29 mm at 0.15 THz). The efficiency can be reduced to 47.7 % for misalignment up to 0.6 mm (almost half of the cross-section dimension). To mitigate the coupling losses, the measured transmission spectra (Fig. 3) are the averaged of two different sets of measurements for both polarizations. Note that each set of measurements includes six different positions (measured in the presence of grating), and their averaged transmission is normalized to the free-space signal (pinhole included). Moreover, some variations can be observed outside the stopbands mostly at low frequencies. The reason for transmission deviations at low frequencies could be that, in simulations, the refractive index of COC is from the fitted data at 0.15 THz frequency.

*Group Velocity Dispersion*
Here, we extract the group-velocity dispersion from measured phase information of short and long gratings. To do this, we calculate the experimental GVD from the second-order derivative of the frequency-dependent propagation constant ($\partial^2\beta/\partial\omega^2$) of the guided modes. Figure 4 presents the experimentally measured (solid lines) and numerically simulated (dashed lines) GVD curves of the short and long gratings. Due the strong index modulation of periodic subwavelength structures, we experimentally achieve maximum of -25 (-25) ps/mm/THz GVD at 0.150 (0.152) THz for short grating [Fig. 4 (a)] and -188 (-88) ps/mm/THz GVD at 0.15 (0.151) THz for long gratings [Fig. 4 (b)] for *x*-pol (*y*-pol.) states. In numerical simulations, we note that both short and long gratings have large negative GVD values of maximum -226 (-186) ps/mm/THz and -340 (300) ps/mm/THz, respectively for *x*-pol (*y*-pol.) waves. We attribute the measured GVD deviations particularly for short grating, to its relatively flat and wide transmission (i.e., the relatively weak coupling), which implies smaller GVDs. Nevertheless, such high negative experimental GVD values have not been reported yet in THz
waveguide gratings, to the best of our knowledge. Interestingly, our long-fabricated grating (Fig. 4b) provides more than 7.5 times larger negative GVD (GVD = -188 ps/mm/THz at 0.15 THz) in less than half of the grating length (43 mm) compared to the THz hollow-core metallic waveguide grating (GVD = -25 ps/mm/THz at 0.14 THz for the grating length of 100 mm with similar transmission losses (-11 dB) of the stopbands [16].

Due to the large negative GVD, our fabricated devices can be useful for dispersion compensation in the THz communication system. For instance, let's consider -60 ps/THz/mm GVD (see the crossings of gray solid lines in Fig. 4 (b)) as a typical value of our fabricated birefringent grating at f=0.148 THz, where the transmission is also relatively high (7.8 dB attenuation compared to pass-band level). Taking 100-km long-haul backup links for satellite assisted wireless communications (generally hundreds of kilometers) as an example here (GVDs of terahertz waves in the atmosphere are $2.5 \times 10^{-5}$ ps/mm/THz at 0.15 THz [16]), the required grating length for zero-GVD compensation is just 41.6 mm — similar to our fabricated long grating sample. Thus, high negative-dispersion in the THz waveguide-based grating can be the key device towards dispersion compensation in future THz communications systems.

Table 2 summarizes the measured characteristics of the waveguide-based grating demonstrations to date in the THz frequency range. First, it shows that our COC-based fiber grating achieves the highest negative GVD for both *x*- and *y*-polarization filtering compared to others. We demonstrate a large negative GVD value of -188 ps/mm/THz (*x*-pol.) in comparison to the GVD value of -25 ps/mm/THz of the hollow-core metallic grating. Moreover, in our case, such high GVD is achieved in less than half the grating length compared to a metallic hollow-core grating length [13]. This shows that our fabricated grating has the advantages of 7.5 times increase in GVD meanwhile in half grating length compared to reported ones, to the best of our knowledge.

**Table 2. Performance comparison of waveguide gratings in the THz region.**

| Attributes | Hollow-core metallic [16] | Notch-type (TOPAS) [25] | Paper-grating [26] | All-silicon grating [23] | COC grating [This paper] | |
|---|---|---|---|---|---|---|
| | | | | | x-pol | y-pol |
| **L (mm)** | 100 | 98 | 70.3 | 8.92 | **43** | **43** |
| **f (THz)** | 0.14 | 0.265 | ~0.369 | 0.275 | **0.15** | **0.152** |
| **ER (dB)** | >10 | ~18 | ~14 | >20 | **8.5** | **7.5** |
| **FWHM (GHz)** | <10 | ~ 4 | 3.5 | 18 | **5** | **5** |
| **GVD (ps/mm/THz)** | ~ -25 | N/A | N/A | N/A | **-188** | **-88** |
| **Polarization** | Polarization independent | Designed for single polarization | Single polarization | Single polarization | **Dual-polarization** | |

L: waveguide grating length, f: filtering frequency, ER: extinction ratio, FWHM: full-width at half-maximum.

Second, the THz grating proposed in this work is the only grating filter that can filter two orthogonal polarizations simultaneously while the others show either single-polarization or no polarization state at all. Interestingly, we also demonstrate the filtering of both polarizations with similar ER and FWHM. Although notch-type dielectric grating [24] and paper grating [25] may have the higher ER, it is possible to improve the ER of our grating by increasing the number of grating cells. For example, our numerical simulations show that ER could increase up to 12 dB in a grating of 59 periods, which is comparable to the ER of paper grating [25]. The FWHM of our long grating is wider than paper or notch type gratings however it is narrow than hollow-core metallic grating. The simulation shows that the minimum FWHM is 5 GHz for our proposed grating, which is reached for the grating period larger than 29. In fact, the moderate filtering frequency range could be beneficial for dispersion compensation, i.e., the design structure would be less sensitive to fabrication imperfections. It should be noted that the reason for larger FWHM of the short length is attributed to weaker modulation strength. Furthermore, we notice that a terahertz grating integrated on effective-medium-clad waveguide [23] has been demonstrated recently, yet it filters only one polarization state.

In conclusion, we have experimentally investigated subwavelength birefringent waveguide-based THz gratings. Two gratings with lengths of 30 mm and 43 mm are fabricated using low-loss COC polymer by micromachining techniques. By taking advantage of the strong index modulation in the subwavelength fiber, we have demonstrated the highest negative GVD (7.5 times large compared to [13]) with the shortest length so far, i.e., 43 mm in comparison to 100 mm. The measurements confirm that the fabricated gratings have negative GVD of -188 ps/mm/THz and -84 ps/mm/THz for *x*- and *y*-polarizations in device length of only 43mm, respectively. We have experimentally characterized the propagation of the two orthogonally polarized guided modes (*x*- and *y*-polarization), emulating the polarization-maintaining nature of the gratings along the propagation direction. Both theoretical and experimental results confirm the filtering of two polarization states simultaneously at 0.15 THz. The proposed compact birefringent grating demonstrates the feasibility of using polymer-based terahertz waveguides for filtering and dispersion compensation in future THz communication systems, particularly using the polarization-division-multiplexing technique to enhance the transmission capacity.

**Funding.** Beijing Natural Science Foundation (4192048), National Natural Science Foundation of China (62075007)

**Acknowledgments.** This work was performed in part at the South Australia node of the Australian National Fabrication Facility. A company established under the National Collaborative Research Infrastructure Strategy to provide nano and microfabrication facilities for Australia's researchers. The authors would like to thank Mr. Qigejian Wang and Mr. Noman Siddique from Shaghik's THz Group at UNSW for assisting with COC material characterization. S. A. acknowledges the support of UNSW Scientia funding.

**Data availability.** Data are available on request from the authors.

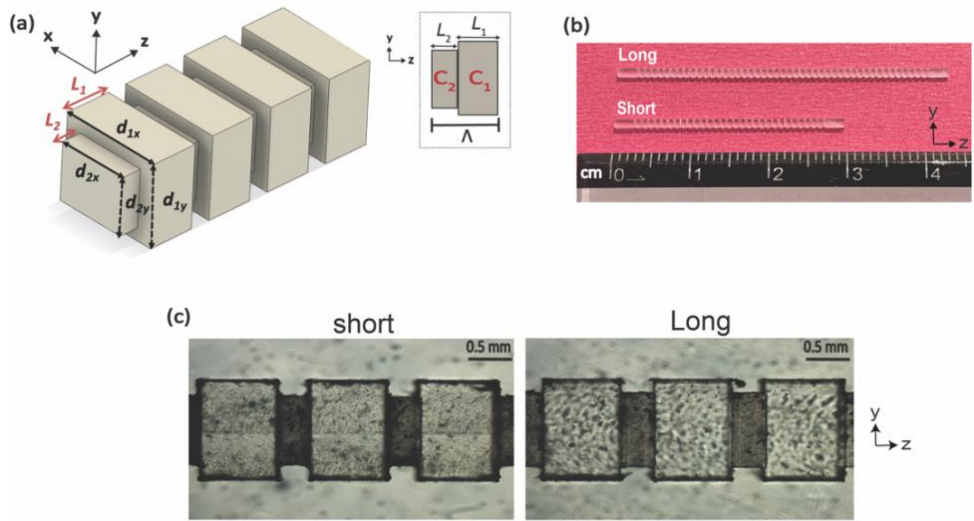

**Fig. 1**. Schematic design and fabricated grating samples. (a) Schematic of multiple unit-cells of the subwavelength birefringent waveguide-based THz grating with the geometrical parameters ($L_1, L_2, d_{1x}, d_{1y}, d_{2x}, d_{2y}$). Inset: The unit cell cross-section consists of the large ($C_1$) and small ($C_2$) sized rectangular cells in the *yz* plane (b) Photographs of the fabricated (COC) grating samples. The number of grating periods is 29 and 45 for short (30-mm) and long (43-mm) gratings, respectively. (c) Microscopic images of the two samples for limited grating units in the *yz* plane.

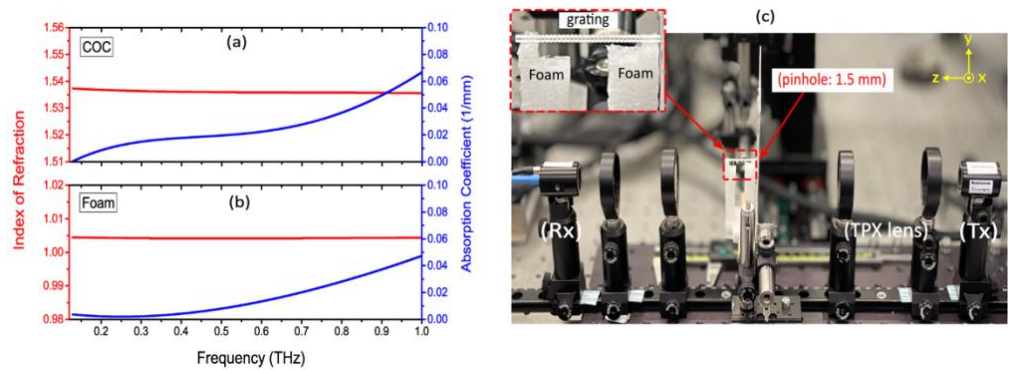

**Fig. 2.** Material characterization and THz-TDS experimental setup. Measured (fitted curve) index of refraction and absorption coefficient of COC (a) and dielectric foam (b) in the frequency range from 0.1 to 1 THz. (c) Image of the fiber-coupled THz-TDS characterization setup. Inset: grating sample on a dielectric foam holder where only the waveguide edges are placed on the foam holder.

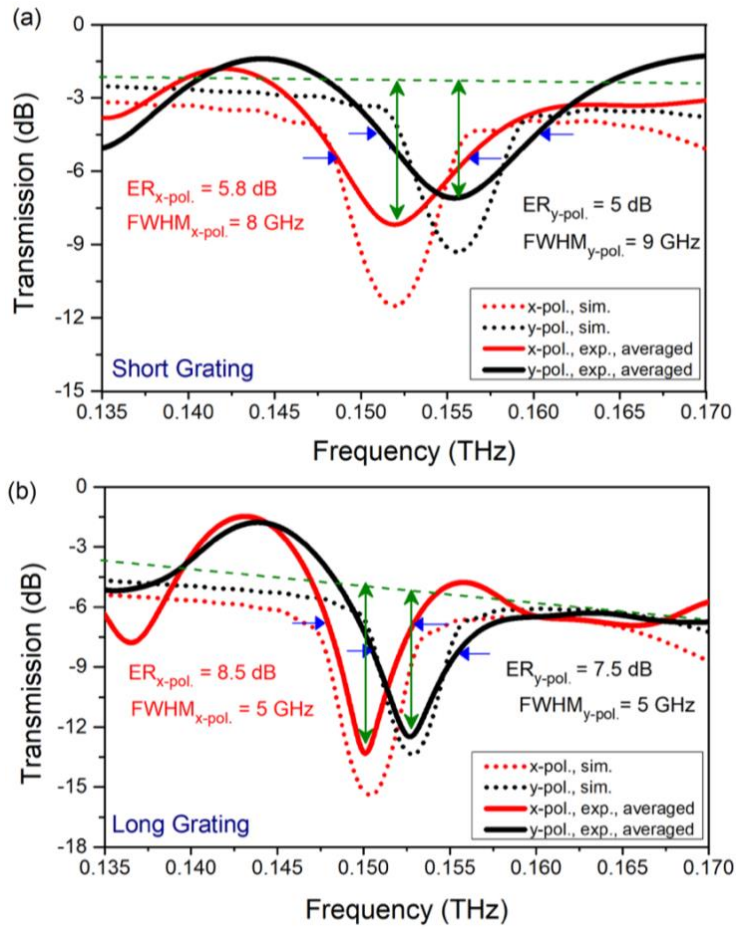

**Fig. 3**. Numerical computation and experimental characterization of THz birefringent fiber grating (a) Short grating (b) Long grating, transmissions for *x*- and *y*-polarizations along with extinction ratios and full-width half-maximums. Red: *x*-polarization, black: *y*-polarization, Dashed-Green: averaged-transmission level outside the stopbands, Solid-Green: extinction ratio measurement, Blue: FWHM measurement. dashed curves: simulation, solid curves: experiment.

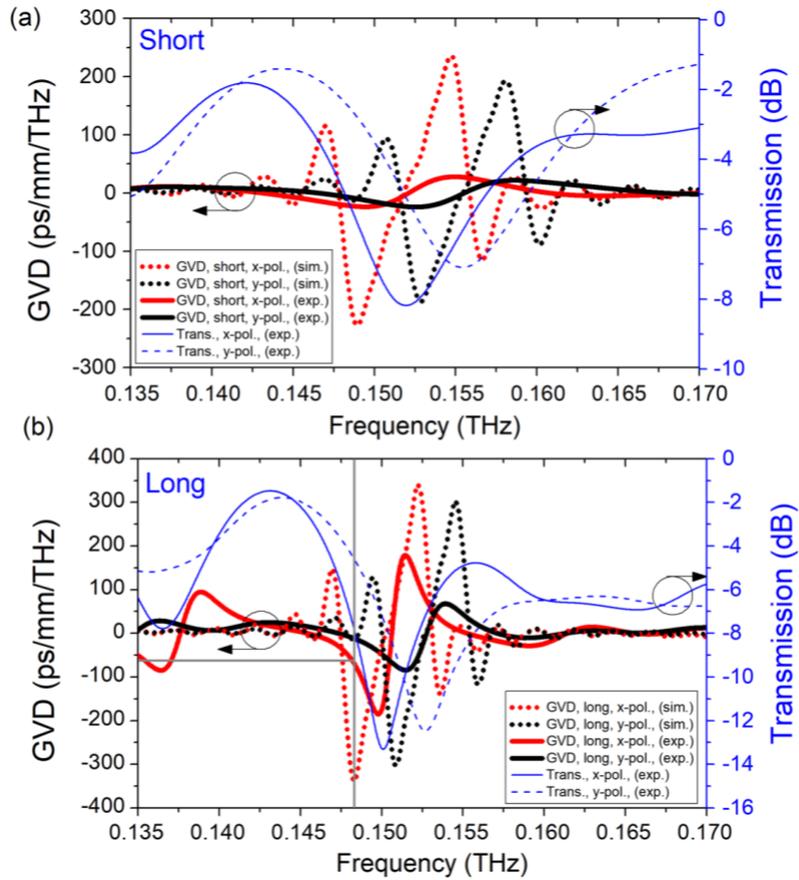

**Fig. 4**. Numerical and experimental group-velocity dispersion comparison as a function of frequency (a) Short grating, (b) Long grating. Red and black curves show *x*-pol. and *y*-pol polarizations, respectively. Note that the crossing of the gray lines in (b) represents the lowest negative GVD value (i.e., -60 ps/mm/THz) of the long grating (*x*-pol.) at relatively high transmission (7.8 dB) in the experimental characterization. Dashed lines: numerical, Solid lines: experiment.